\documentclass[11pt,a4paper,final]{article}
\usepackage[latin1]{inputenc}
\usepackage{times}
\usepackage{a4wide}
\usepackage{amsfonts}
\usepackage{amssymb}
\usepackage{amsmath}
\usepackage[notref,notcite]{showkeys}
\usepackage{ifpdf}
\ifpdf
\usepackage[pdftex]{graphicx}
\usepackage[pdftex,unicode,implicit]{hyperref}
\hypersetup{%
  pdftitle    = {Non-supersymmetric Black Holes of $N=2,d=5$ Supergravity},
  pdfkeywords = {supergravity, black hole}
  pdfauthor   = {P. Meessen and  T. Ortín},
  pdfcreator  = {pdf\LaTeXe\ with package \flqq hyperref\frqq},
  pdfproducer = {pdf\LaTeXe\ with package \flqq hyperref\frqq},
  pdfpagemode = None,  
  pdffitwindow= true,  
  unicode     = true,
  plainpages  = true,
  colorlinks  = true,  
  citecolor   = blue,
  urlcolor    = red,
  linkcolor   = black
}
\newcommand{\hepth}[1]{{\tt
\href{http://www.arXiv.org/abs/hep-th/#1}{hep-th/#1}}}

\newcommand{\arxiv}[1]{{\tt
\href{http://www.arXiv.org/abs/#1}{arXiv:#1}}}
\else
  \usepackage[dvips]{graphicx}
  \usepackage[unicode,implicit]{hyperref}
  \newcommand{\hepth}[1]{{\tt hep-th/#1}}

  \newcommand{\arxiv}[1]{{\tt arXiv:#1}}
\fi
\usepackage{tikz}
\newcommand{\FPAUO}[2]{
\tikz[scale=.13,
         Uniovi/.style={color=gray, fill=gray}
 ] {
 \fill[Uniovi] (0,0) circle (10);
 \fill[white] (0,7) circle (1.5);
 \draw[Uniovi] (-2,7.5) rectangle (2,5.5);
 \fill[white] (-0.3,6.6) rectangle (0.3,0);   
 \fill[white] ( -0.9,6.2) rectangle (.9 ,5.6);
 \fill[white] (-1.4, 5.2) rectangle (1.4, 4.6);
 \fill[white] (0,0) ellipse (3.5 and 4);
 \fill[Uniovi] (-2.5,0.3) rectangle (2.5,-0.3);
 \fill[Uniovi] (-2,2.3) rectangle (2,1.7);
 \fill[Uniovi] (-2,-2.3) rectangle (2,-1.7);
 \fill[white] (-4.5,5.5) rectangle (-2.7,4.9);
 \fill[white] (-3.9,6.1) rectangle (-3.3,4.3);
 \fill[white] (4.5,5.5) rectangle (2.7,4.9);
 \fill[white] (3.9,6.1) rectangle (3.3,4.3);
 \foreach \x in { 0,..., 3 }
   \foreach \y in { 0,...,\x}
    {
     \fill[white] (-6-\x*0.7+\y*1.4,3.5-\x *1.97) -- (-5.6-\x*0.7+\y*1.4,2.4-\x *1.97) -- (-6.4-\x*0.7+\y*1.4,2.4-\x *1.97) -- cycle;
     \fill[white] (6-\x*0.7+\y*1.4,3.5-\x *1.97) -- (5.6-\x*0.7+\y*1.4,2.4-\x *1.97) -- (6.4-\x*0.7+\y*1.4,2.4-\x *1.97) -- cycle;
   };
 \draw (0,-6) node[
                               text centered, 
                               color=white, 
                               font={\fontsize{8}{4}\sffamily\selectfont}
                             ] {FPAUO-#1/#2};
}} 
\makeatletter
\@addtoreset{equation}{section}
\makeatother

\begin{document}

\begin{flushright}
\small
\FPAUO{11}{11}\\
IFT-UAM/CSIC-11-55\\  
July  $27^{\rm th}$, 2011\\
\normalsize
\end{flushright}

\begin{center}

\vspace{1cm}

{\LARGE {\bf Non-Extremal Black Holes\\[.5cm] of $N=2,d=5$ Supergravity}}

\vspace{.5cm}

\begin{center}

{\sl\large Patrick Meessen$^{\aleph}$}
\footnote{E-mail: {\tt meessenpatrick@uniovi.es}},
{\sl\large and T.~Ort\'{\i}n$^{\diamond}$}
\footnote{E-mail: {\tt Tomas.Ortin@csic.es}}

\vspace{.5cm}

${}^{\aleph}${\it HEP Theory Group, Departamento de F\'{\i}sica, Universidad de Oviedo\\ 
        Avda.~Calvo Sotelo s/n, E-33007 Oviedo, Spain}\\

\vspace{.6cm}

${}^{\diamond}${\it Instituto de F\'{\i}sica Te\'orica UAM/CSIC\\
C/ Nicol\'as Cabrera, 13-15,  C.U.~Cantoblanco,  E-28049-Madrid, Spain}\\

\end{center}

\vspace{.5cm}

{\bf Abstract}

\begin{quotation}

  {\small 
    We study the generalization of the Ansatz of Galli {\em et al.\/} \cite{Galli:2011fq} for
    non-extremal black holes of $N=2,d=4$ supergravities 
    for a simple model of $N=2,d=5$ supergravity with a vector multiplet whose
    moduli space has two branches. We use the formalism of Ferrara, Gibbons
    and Kallosh \cite{Ferrara:1997tw}, which we generalize to any dimension
    $d$. We find that the equations of motion of the model studied can be
    completely integrated without the use of our Anstaz (which is,
    nevertheless, recovered in the integration). The family of solutions found
    (common to both branches) is characterized by five independent
    parameters: the mass $M$, the electric charges $q_{0},q_{1}$, the
    asymptotic value of the scalar at infinity $\phi_{\infty}$ and the scalar
    charge $\Sigma$. The solutions have a singular horizon whenever $\Sigma$
    differs from a specific expression
    $\Sigma_{0}(M,q_{0},q_{1},\phi_{\infty})$ ({\em i.e.\/} when there is
    primary scalar hair $\Sigma-\Sigma_{0}\neq 0$). The family of regular
    black holes interpolates between its two extremal limits. The
    supersymmetry properties of the extremal solutions depend on the choice of
    branch: one is always supersymmetric and the other non-supersymmetric in
    one branch and the reverse in the other one.
}

\end{quotation}

\end{center}

\newpage
\pagestyle{plain}
\section*{Introduction}
In a recent paper \cite{Galli:2011fq} Galli~\textit{et al.} proposed a
general Ansatz to find non-extremal black-hole solutions of $N=2,d=4$
supergravity theories coupled to vector multiplets, that makes crucial use of
the formalism developed by Ferrara, Gibbons and Kallosh (FGK) in
ref.~\cite{Ferrara:1997tw}.\footnote{
  For previous work on near-extremal and
  non-extremal solutions see
  {\em e.g.\/} refs.~\cite{Horowitz:1996fn,Cvetic:1996kv,Ortin:1996bz,Kastor:1997wi,Behrndt:1997as,Cardoso:2008gm,Gruss:2009kr,Gruss:2009wm,Mohaupt:2010fk}.
}
The Ansatz consists of a systematic deformation of the corresponding
supersymmetric (extremal) solutions to the same model which has to be plugged
into the equations of motion derived by FGK to determine the values of the
integration constants, something that needs to be done for each particular
model. 

The Ansatz can be generalized to higher dimensions by using the
corresponding generalization of the FGK formalism, but it may only work for
$N=2$-type theories for which the metric functions of supersymmetric solutions
are homogenous of a certain degree in harmonic seed functions. In this paper we
want to study a generalization of ref.~\cite{Galli:2011fq}'s Ansatz
for the $N=2,d=5$ supergravity case, and we will generalize the FGK formalism
and the results obtained in refs.~\cite{Gibbons:1996af,Ferrara:1997tw} to
arbitrary dimensions. We will then construct the non-extremal black-hole
solutions of a simple model of $N=2,d=5$ supergravity with just one vector
multiplet (and, therefore, one scalar field).

This paper is organized as follows: Section~\ref{sec-FGK} is devoted to the
generalization of the results of \cite{Ferrara:1997tw} to $d\geq 4$
dimensions.  In Section~\ref{sec-FGKn2d5} we adapt the results of the previous
section to the particular case of $N=2,d=5$ theories with vector multiplets.
In Section~\ref{sec-model} we construct the general black-hole solutions of a
simple model of $N=2,d=5$ supergravity, studying first the supersymmetric
ones, which can be constructed using well-known recipes.
Section~\ref{sec-conclusions} contains our conclusions.
\section{The FGK formalism in $d\geq 4$}
\label{sec-FGK}
In order to generalize the results of ref.~\cite{Ferrara:1997tw} to $d\geq 4$
we first need to find a suitable generalization of their radial coordinate $\tau$, a goal
that can be achieved relatively easily:\footnote{
   Observe that the case $d=5$ was treated in ref.~\cite{Mohaupt:2010fk}.
}
consider the $d$-dimensional non-extremal Reissner-Nordstr\"om
(RN) family of solutions. If we normalize the $d$-dimensional Einstein-Maxwell
action as (see {\em e.g.\/} ref.~\cite{kn:PRC})
\begin{equation}
\label{eq:d-dimensionalEinsteinMaxwell}
\mathcal{I}[g_{\mu\nu},A_{\mu}]
= \frac{1}{16\pi G^{(d)}_{N}} {\displaystyle\int} d^{d}x \sqrt{|g|}\
\left[R -\tfrac{1}{4}F^{2}\right]
\, ,
\end{equation}

\noindent
where $G^{(d)}_{N}$ is the $d$-dimensional Newton constant. Then, the metric can be
put in the form
\begin{eqnarray}
  \label{eq:RNd}
ds^{2} & = & H^{-2}W dt^{2} - H^{\frac{2}{d-3}}\left[W^{-1}dr^{2}
  +r^{2}d\Omega^{2}_{(d-2)} \right]\, , \\
 & & \nonumber \\
\label{eq:RNd.1}
H & = & {\displaystyle 1 +\frac{h}{r^{d-3}}\, , \hspace{1cm}
W = 1 -\frac{2\mathcal{B}}{r^{d-3}}}\, ,  
\end{eqnarray}

\noindent
where $d\Omega^{2}_{(d-2)}$ is the metric of the unit $(d-2)$-sphere and the
constant $h$ and the non-extremality parameter $\mathcal{B}$ are given
by\footnote{
   In $d=4$, $\mathcal{B}$ is usually called $r_{0}$ or $c$.
}
\begin{equation}
\mathcal{B} = \frac{4\pi G^{(d)}_{N}}{(d-2)\omega_{(d-2)}} \sqrt{M^{2}
  -2\frac{(d-2)}{(d-3)}q^{2}}\, , 
\hspace{1cm}
h = \frac{4\pi G^{(d)}_{N}}{(d-2)\omega_{(d-2)}}M -\mathcal{B}\, .  
\end{equation}

\noindent
In the above expressions $\omega_{(d-2)}$ is the volume of the unit $(d-2)$-sphere,
$M$ is the ADM mass and $q$ the canonically-normalized electric
charge.

The metric (\ref{eq:RNd}) describes the exterior of a RN black hole with the (outer) event
horizon being located at $r^{d-3}=2\mathcal{B} \geq 0$. The (inner) Cauchy horizon would, in
principle, be located at $r^{d-3}=-h \leq 0$: this corresponds to a real value of $r$
only for even $d$; for odd $d$, the Cauchy horizon is not covered by these coordinates.  

When $\mathcal{B}=0$ the function $W$ effectively disappears from the metric and we recover
the extremal RN black hole in isotropic coordinates. As is well-known, in this
limit there are many other solutions of the same form with $H$ replaced by an
arbitrary function harmonic on Euclidean $\mathbb{R}^{d-1}$. In this sense,
the above non-extremal metric can be understood as a deformation of the
extremal one by an additional \textit{harmonic} function $W$ (called
\textit{Schwarzschild} or \textit{non-extremality factor}) containing the (non-)BPS
parameter $\mathcal{B}$. This kind of deformations have been used to find
non-extremal solutions in {\em e.g.\/} refs.~\cite{Ortin:1996bz,Mohaupt:2010fk}.\footnote{
 As one can see from ref.~\cite{Mohaupt:2010fk} the solution for non-extreme black holes
 that we are going to construct, can, due to the special properties of supersymmetric couplings, 
 be coordinate-transformed to a solution with a Schwarzschild factor.
}

If we perform the coordinate transformation 
\begin{equation}
r^{d-3} = \frac{2\mathcal{B}}{1-e^{-2\mathcal{B} \rho}}\, ,  
\end{equation}

\noindent
in the above metric we find that it takes the conformastatic form 
\begin{equation}
\label{eq:conformastatic}
  ds^{2} = e^{2U}dt^{2} - e^{-\frac{2}{d-3}U}\gamma_{\underline{m}\,
    \underline{n}}dx^{\underline{m}} dx^{\underline{m}}\, , 
\end{equation}

\noindent
where the function $e^{2U}$ is given by
\begin{equation}
  \label{eq:e2U}
  e^{2U}  = \hat{H}^{-2} e^{-2\mathcal{B}\rho} \hspace{.5cm}\mbox{with}\hspace{.5cm}
  \hat{H} \ =\ 
     \frac{h+2\mathcal{B}}{2\mathcal{B}} -\frac{h}{2\mathcal{B}} e^{-2\mathcal{B}\rho}\, , 
\end{equation}

\noindent
and the spatial background metric, $\gamma$, is given by 
\begin{equation}
\label{eq:background}
\gamma_{\underline{m}\,\underline{n}}dx^{\underline{m}} dx^{\underline{m}}
=
\left[\frac{\mathcal{B}}{\sinh{(\mathcal{B}\rho)}} \right]^{\frac{2}{d-3}}  
\left[
\left(\frac{\mathcal{B}}{\sinh{(\mathcal{B}\rho)}} \right)^{2}
\frac{d\rho^{2}}{(d-3)^{2}}
+d\Omega^{2}_{(d-2)}\right]\, .
\end{equation}

The coordinate $\rho$ is the higher-dimensional generalization of the $\tau$ of
ref.~\cite{Ferrara:1997tw} we were looking for. In fact, in $d=4$ their relation is
$\rho=-\tau$. The main difference with $\tau$ is that the event horizon is at
$\rho \rightarrow +\infty$ instead of $-\infty$; furthermore, the Cauchy horizon, which
in $d=4$ could be reached at $\tau\rightarrow +\infty$, is not covered by
$\rho$ because, in general, it cannot take negative values due to the
fractional power in $\gamma$. 
In the extremal limit, {\em i.e.\/} when $\mathcal{B}\rightarrow 0$, the background metric takes the form 
\begin{equation}
\label{eq:extremalmetric}
\gamma_{\underline{m}\,\underline{n}}dx^{\underline{m}} dx^{\underline{m}}
=
\frac{1}{\rho^{\frac{2}{d-3}}}
\left[
\left(\frac{d\rho}{(d-3) \rho}\right)^{2}
+d\Omega^{2}_{(d-2)}\right]\, , 
\end{equation}

\noindent
which is nothing but the Euclidean metric on $\mathbb{R}^{d-1}$ as can be seen
by the coordinate change $\rho = r^{3-d}$;
%
needless to say, in the limit $\mathcal{B}=0$ the function $\hat{H}$ becomes a harmonic function
on $\mathbb{R}^{d-1}$.

It is reasonable to expect that all static black-hole metrics in $d\geq 4$
dimensions can be brought to the conformastatic form
eq.~(\ref{eq:conformastatic}) with the background metric (\ref{eq:background}).
In the next section we will also assume that the
metric function $e^{-2U}$ of the non-extremal black holes of $N=2,d=5$
supergravity can be obtained from that of the extremal ones by replacing the
harmonic functions $H_{I}$ by hatted harmonic functions of the form
$\hat{H}_{I}=a_{I}+b_{I}e^{-\mathcal{B}\rho}$ and adding an overall factor of
$e^{\mathcal{B}\rho}$ as in eq.~(\ref{eq:e2U}).

Let us consider now the $d$-dimensional action
\begin{equation}
\mathcal{I}[g_{\mu\nu},A^{\Lambda}{}_{\mu},\phi^{i}] = \int d^{d}x 
\left\{
R  +\mathcal{G}_{ij}\partial_{\mu}\phi^{i}\partial^{\mu}\phi^{j} 
+2I_{\Lambda\Sigma}F^{\Lambda}{}_{\mu\nu}F^{\Sigma\, \mu\nu}
\right\}\, ,
\end{equation}

\noindent
where the $I_{\Lambda\Sigma}$ are given functions of the scalars
$\phi^{i}$ and are supposed to form an invertible, negative definite matrix. 

In $d>4$ dimensions there could be higher-rank potentials in the action, but
they should not couple to black holes. Of course, their consistent truncation
from the action could place additional constraints on the remaining fields,
but this analysis has to be made on a case by case basis and one could always
impose those constraints on the solutions to the above unconstrained action. In
odd dimensions there could also be Chern-Simons terms for the 1-forms
$A^{\Lambda}{}_{\mu}$. However, those terms will only contribute to the
equations of motion when we consider objects magnetically charged with respect
to the 1-forms, {\em i.e.\/} electrically charged with respect to their dual
$(d-3)$-forms, but these would not be black holes in $d>4$.  Therefore, we can
conclude that the above action is general enough to cover all or most of the
possible (necessarily electrically) charged black-hole solutions in $d>4$. In
$d=4$ there is an additional term involving only scalars and 1-forms related
to the fact that only in $d=4$ dimensions black holes can have magnetic
charges on top of the electric ones.

Plugging the assumptions of time-independence of all fields and a metric of the form
eqs.~(\ref{eq:conformastatic},\ref{eq:background}) into the equations of motion
resulting from the action, and using the
conservation of the electric charges $q_{\Lambda}$,
we are left with a reduced system of differential equations in
$\rho$ that can be derived from the so-called {\em geodesic action}
\begin{equation}
\label{eq:Effe40}
\mathcal{I}[U,\phi^{i}] 
= \int d\rho 
\left\{
(\dot{U})^{2}
+\textstyle{(d-3)\over (d-2)}\ \mathcal{G}_{ij}\dot{\phi}^{i}\dot{\phi}^{j}
-e^{2U}V_{\rm bh}
\right\}\, ,   
\end{equation}

\noindent
where the black-hole potential is given by 
\begin{equation}
V_{\rm bh} \, =\, \alpha^{2}\ \textstyle{2(d-3)\over (d-2)}\ 
I^{\Lambda\Sigma}q_{\Lambda}q_{\Sigma}\, ,  
\end{equation}

\noindent
$\alpha$ being a constant related to the normalization of the charge to be
determined later;
one also finds a relation between the Hamiltonian corresponding to the action (\ref{eq:Effe40})
and the non-extremality parameter $\mathcal{B}$, namely
\begin{equation}
\label{eq:HamBsquare}
  (\dot{U})^{2}\, +\,
  \textstyle{(d-3)\over (d-2)}\ \mathcal{G}_{ij}\dot{\phi}^{i}\dot{\phi}^{j}
  \, +\, e^{2U}V_{\rm bh}   
  \, =\, \mathcal{B}^{2}\, .
\end{equation}

Assuming regularity of the fields at the horizon, it is possible to derive
generalizations of the theorems proven in ref.~\cite{Ferrara:1997tw}: for
extremal black holes, in the $\rho\rightarrow +\infty$ limit
\begin{equation}
e^{U} \; \sim\; \frac{1}{\rho}\,
\left[ \frac{A_{\rm h}}{\omega_{(d-2)}}
\right]^{-\frac{(d-3)}{(d-2)}} \; ,
\end{equation}

\noindent
where $A_{\rm h}$ is the area of the event horizon. Furthermore, this area is
given by 
\begin{equation}
A_{\rm h} = \omega_{(d-2)} 
\left[- V_{\rm bh}(\phi^{i}_{\rm h}) \right]^{\frac{(d-2)}{2(d-3)}}\, ,  
\end{equation}

\noindent
where the values of the scalars at the horizon, $\phi^{i}_{\rm h}$, extremize
the black-hole potential
\begin{equation}
\left. \partial_{i}V_{\rm bh}\right|_{\phi^{i}_{\rm h}}=0\, .
\end{equation}

For general (extremal or non-extremal) black holes, defining the mass $M$ and
the scalar charges $\Sigma^{i}$ by the asymptotic ({\em i.e.\/} $\rho \rightarrow 0$)
behavior of the metric function and scalars as
\begin{equation}
  \label{eq:1}
  U \ \sim\ -M\rho \hspace{.5cm},\hspace{.5cm} \phi^{x}\ \sim\ \phi^{x}_{\infty}\ -\ \Sigma^{x}\ \rho \; ,
\end{equation}

\noindent
we find from eq.~(\ref{eq:HamBsquare})
\begin{equation}
\label{eq:BBound}
M^{2} \ +\ \textstyle{(d-3)\over (d-2)}\ \mathcal{G}_{ij}(\phi_{\infty}) \Sigma^{i}\Sigma^{j}
\ +\ V_{\rm bh}(\phi_{\infty},q) \, =\, \mathcal{B}^{2}\, .  
\end{equation}

Finally, the entropy $S = A_{\rm h}/(4G^{(d)}_{N})$ and temperature,
$T$, of the black-hole event horizon are related to the
non-extremality parameter by generalization of the Smarr formula
\cite{Gibbons:1996af}
\begin{equation}
\mathcal{B} \, =\, \frac{16\pi G^{(d)}_{N}}{(d-3)\omega_{(d-2)}}\ ST\; .  
\end{equation}

Observe that the mass $M$ defined above is identically to the ADM mass if we
set 
\begin{equation}
\frac{8\pi G^{(d)}_{N}}{(d-2)\omega_{(d-2)}} =1\, ,  
\end{equation}

\noindent
as we will do from now on,\footnote{
  With this choice, to have canonically-normalized
  charges in the black-hole potential $\alpha$ must take the value
  \begin{equation}
     \alpha= \frac{(d-2)}{\sqrt{2}(d-3)}\, .  
  \end{equation}
  This is not the most convenient normalization, though, because, with it, the
  relation between mass and charge of an extremal RN black hole is 
  \begin{equation} 
     M^{2} = \textstyle{2(d-2)\over (d-3)}\ q^{2}\, ,  
  \end{equation}
  and we will choose a different one in the next section.
}
we have
\begin{equation}
\label{eq:SSmarr}
S= \frac{2\pi}{(d-2)\omega_{(d-2)}} A_{\rm h}
\hspace{1cm}\mbox{whence}\;\;\;\;
ST = \textstyle{(d-3)\over 2(d-2)}\ \mathcal{B}\, . 
\end{equation}
\subsection{The FGK formalism for $N=2,d=5$ theories}
\label{sec-FGKn2d5}
The relevant part of the bosonic action of $N=2,d=5$ supergravity theories
coupled to $n$ vector multiplets is, using the conventions of
refs.~\cite{Bellorin:2006yr,Bergshoeff:2004kh},
\begin{equation}
\mathcal{I}[g_{\mu\nu},A^{I}{}_{\mu},\phi^{x}] = \int d^{5}x 
\left\{
R  +\tfrac{1}{2}g_{xy}\partial_{\mu}\phi^{x}\partial^{\mu}\phi^{y} 
-\tfrac{1}{8}a_{IJ}F^{I}{}_{\mu\nu}F^{J\, \mu\nu}
\right\}\, ,
\end{equation}

\noindent
where $I,J=0,1,\cdots,n$ and $x,y=1,\cdots,n$. The scalar target spaces are
determined by the existence of $n+1$ functions $h^{I}(\phi)$ of the $n$
physical scalar subject to the constraint
\begin{equation}
C_{IJK}h^{I}h^{J}h^{K}=1\, ,  
\end{equation}

\noindent
where $C_{IJK}$ is a completely symmetric constant tensor that determines the
model. Defining
\begin{equation}
h_{I} \equiv C_{IJK}h^{J}h^{K}\hspace{1cm}\mbox{(whence $h_{I}h^{I}\ =\ 1$)}
\end{equation}

\noindent
the matrix $a_{IJ}$ can be expressed as
\begin{equation}
a_{IJ} = -2C_{IJK}h^{K}+3h_{I}h_{J}\, ,  
\end{equation}

\noindent
and can be used to raise and lower the index of the functions $h^{I}$. We also define
\begin{equation}
  \label{eq:9}
  h^{I}{}_{x} \ \equiv\ -\sqrt{3} \partial_{x}h^{I}\;\; ,\;\;
  h_{I\, x} \ \equiv\ a_{IJ}h^{J} \ =\ +\sqrt{3} \partial_{x}h_{I}\; ,
\end{equation}

\noindent
which are orthogonal to the $h^{I}$ with respect to the metric
$a_{IJ}$. Finally, the target-space metric is given by
\begin{equation}
\label{eq:a-1}
g_{xy}\  \equiv\  a_{IJ}h^{I}{}_{x} h^{J}{}_{y} \;\;
\xrightarrow{\ \mbox{which implies}\ }
\;\;
a^{IJ} = h^{I}h^{J} +g^{xy}h^{I}{}_{x}h^{J}{}_{y}\, .
\end{equation}

Adapting the results of the previous section to these conventions and
definitions we get the effective action
\begin{equation}
\label{eq:effectived5}
\mathcal{I}[U,\phi^{x}] 
= \int d\rho 
\left\{
(\dot{U})^{2}
+\tfrac{1}{3}g_{xy}\dot{\phi}^{x}\dot{\phi}^{y}
-e^{2U}V_{\rm bh}
\right\}\, ,   
\end{equation}

\noindent
and Hamiltonian constraint (\ref{eq:HamBsquare}) becomes
\begin{equation}
\label{eq:Effe2a}
(\dot{U})^{2}
+\tfrac{1}{3}g_{xy}\dot{\phi}^{x}\dot{\phi}^{y}
+e^{2U}V_{\rm bh}
= 
\mathcal{B}^{2}\, ,  
\end{equation}

\noindent
where the black-hole potential with the choice of normalization
$\alpha^{2}=3/32$, is given by
\begin{equation}
-V_{\rm bh} \ =\ a^{IJ}q_{I}q_{J} 
                \ =\  \mathcal{Z}^{2}
                   +3g^{xy}\partial_{x}\mathcal{Z}\partial_{y}\mathcal{Z}\, ,
\end{equation}

\noindent
where we defined the {\em central charge} $\mathcal{Z}(\phi,q)\equiv h^{I}q_{I}$
and used eq.~(\ref{eq:a-1}) in order to obtain the last expression.
The supersymmetric black holes of these theories satisfy 
\begin{equation}
\label{eq:Effe1a}
\left. \partial_{x}\mathcal{Z}\right|_{\rm \phi_{\rm h}}\ =\ 0\hspace{.3cm}
\xrightarrow{\;\;\; \mbox{whence}\;\;\; }\hspace{.3cm}
\left. \partial_{x}V_{\rm bh}\right|_{\rm \phi_{\rm h}}=0\, ,
\end{equation}
{\em i.e.\/} the values the physical scalar fields take at the horizon extremize the central 
charge and the black hole potential;
in fact, all extremal black-hole solutions of the theory satisfy the latter equation
but only the supersymmetric ones satisfy also the former. Furthermore, the
supersymmetric ones saturate the BPS bound
\begin{equation}
M = \mathcal{Z}(\phi_{\infty},q)\, .  
\end{equation}
The supersymmetric, and therefore extremal, black-hole solutions
\cite{Gauntlett:2002nw,Gutowski:2004yv,Gutowski:2005id} are completely
determined by $n+1$ real harmonic functions on Euclidean $\mathbb{R}^{4}$
\begin{equation}
\label{eq:harmonicfunctions}
I_{I} = I_{I\, \infty} +q_{I}\rho\, .  
\end{equation}

\noindent
The fields of the supersymmetric solutions are related to these function by
\begin{equation}
\label{eq:stabil}
e^{-U}\ h_{I}(\phi ) \, =\, I_{I}\, .  
\end{equation}

\noindent
These equations must be solved for $U=U_{susy}(I)$ and the physical scalars $\phi^{x} =\phi^{x}_{susy}(I)$
using the constraints of real special geometry. 

Galli {\em et al.\/}'s Ansatz \cite{Galli:2011fq} for the non-extremal black-holes solutions is a deformation of
the supersymmetric extremal solutions $U_{\rm susy}(I),\phi_{\rm susy}^{x}(I)$, namely
\begin{equation}
  \label{eq:10}
  U \ =\ U_{\rm susy}(\hat{I}) -2\mathcal{B}\ \rho \hspace{.3cm},\hspace{.3cm}
  \phi^{x} \ =\ \phi^{x}_{\rm susy}(\hat{I})\, , 
\end{equation}

\noindent
where the functions $\hat{I}_{I}$ have the form
\begin{equation}
\hat{I}_{I} \, =\, a_{I} +b_{I}e^{-2\mathcal{B}\rho}\, .  
\end{equation}
%
\section{A simple model of $N=2$, $d=5$ supergravity and its black holes}
\label{sec-model}
Let us consider a simple model with one vector multiplet determined by
$C_{011}=1/3$;\footnote{
  This model can be obtained by dimensional reduction of minimal $d=6$ $N=(1,0)$ supergravity.
}
in terms of the physical, unconstrained, scalar $\phi$ we find
that the model has two branches, labeled by $\sigma=\pm 1$:
\begin{equation}
  \begin{array}{rclrcl}
h_{(\sigma )}^{0} & = & e^{\sqrt{\frac{2}{3}}\phi}\, ,
\hspace{1cm}& 
h_{(\sigma )}^{1} & = &  \sigma\ e^{-\frac{1}{\sqrt{6}}\phi}\, , \\ 
& & & & & \\
h_{(\sigma )\, 0} & = & \tfrac{1}{3} e^{-\sqrt{\frac{2}{3}}\phi}\, ,
\hspace{1cm}& 
h_{(\sigma )\, 1} & = & \tfrac{2}{3} \sigma\ e^{\frac{1}{\sqrt{6}}\phi}\, . \\ 
\end{array}
\end{equation}

\noindent
The scalar metric $g_{\phi\phi}$ and the vector field strengths metric
$a_{IJ}$ take exactly the same values in both branches:
\begin{equation}
  \label{eq:11}
  g_{\phi\phi} \ =\ 1 \hspace{.5cm},\hspace{.5cm}
  a_{IJ} \ =\ \tfrac{1}{3}\left(
        \begin{array}{cc}
              e^{-2\sqrt{\frac{2}{3}}\phi} & 0 \\
              0 &   e^{\sqrt{\frac{2}{3}}\phi} \\
        \end{array}
        \right)\, ,
\end{equation}

\noindent
and, therefore, the bosonic parts of both models and their classical
solutions are identical. Since the functions $h_{(\sigma )}^{I}(\phi)$ differ,
the fermionic structure will be different. In particular, the central charge
in the $\sigma$-branch is
\begin{equation}
\mathcal{Z}_{(\sigma )}  = q_{0}e^{\sqrt{\frac{2}{3}}\phi}
+ \sigma q_{1}e^{-\frac{1}{\sqrt{6}}\phi}\, .  
\end{equation}

The  black-hole potential, being a property of the bosonic part of the
theory, is identical in both branches:
\begin{equation}
-V_{\rm bh}  = 
\tfrac{3}{2}\left[   
2q^{2}_{0}e^{2\sqrt{\frac{2}{3}}\phi}
+q^{2}_{1}e^{-\sqrt{\frac{2}{3}}\phi}
\right]\, .
\end{equation}

The black-hole potential is extremized for 
\begin{equation}
\phi_{\rm h} = 
-\sqrt{\tfrac{2}{3}} \log{\left(\pm \sigma \frac{2q_{0}}{q_{1}}\right)}\, .  
\end{equation}

Since $\pm\sigma 2q_{0}/q_{1} >0$, the upper sign (which corresponds to the
supersymmetric case in the $\sigma$-branch, as it extremizes the central
charge) requires the following relation between the signs $s_{I}(\equiv q_{I}/|q_{I}|$ of the
charges $q_{I}$
\begin{equation}
s_{0} = \sigma s_{1}\, ,  
\end{equation}

\noindent
while the lower one (non-supersymmetric in the $\sigma$-branch) requires
\begin{equation}
s_{0} = -\sigma s_{1}\, .
\end{equation}

The same bosonic solution will be supersymmetric in the $\sigma$-branch and
non-supersymmetric in the $(-\sigma)$-branch.
We are going to construct the supersymmetric solutions of the $\sigma$-branch
next; the non-supersymmetric solutions of the $(-\sigma )$-branch will be constructed
at the same time.
\subsection{Supersymmetric and non-supersymmetric extremal solutions}
\label{sec-susysol}
According to the general prescription, the extremal solutions are given by two real harmonic
functions of the form eq.~(\ref{eq:harmonicfunctions}), and are related to $U$
and $\phi$ by eqs.~(\ref{eq:stabil}), which in this case take the form
\begin{equation}
  \label{eq:2}
  I_{0} \ =\ \tfrac{1}{3}e^{-U_{\rm susy}}\ e^{-\sqrt{\frac{2}{3}}\phi_{\rm susy}} \hspace{.5cm},\hspace{.5cm}
  I_{1} \ =\ \tfrac{2}{3}\ \sigma\ e^{-U_{\rm susy}}\ e^{\frac{1}{\sqrt{6}}\phi_{\rm susy}} \; .
\end{equation}

\noindent
Solving for $U_{\rm susy}$ and $\phi_{\rm susy}$ we get
\begin{equation}
  \label{eq:USUSY}
  e^{-U_{\rm susy}} \ =\ \left(\tfrac{3^{3}}{2^{2}} I_{0}I^{2}_{1} \right)^{1/3} \hspace{.4cm},\hspace{.4cm}
  \phi_{\rm susy} \ =\  -\sqrt{\tfrac{2}{3}} \log\left(\sigma \frac{2I_{0}}{I_{1}}\right)\; .
\end{equation}

\noindent
The regularity and well-definedness of these fields impose some restrictions
on the harmonic functions, to wit
\begin{enumerate}
\item[i)] They should not vanish at any finite value of $\rho$: this requirement
  relates the signs of $q_{I}$ and $I_{I}$.
\item[ii)] $\mathrm{sign}(I_{0})=\sigma\ \mathrm{sign}( I_{1})$ everywhere for
  $\phi_{\rm susy}$ to be well-defined in the $\sigma$-branch. This implies,
  in particular, that $s_{0}=\sigma s_{1}$ which is the relation we found for the
  supersymmetric critical points. There are therefore for each branch two
  supersymmetric cases which are disjoint in charge space: $s_{0}=+1,
  s_{1}=\sigma$ and $s_{0}=-1, s_{1}=-\sigma$.
\item[iii)] For $U_{\rm susy}$ to be well-defined ($e^{-U}>0$) only $I_{0}>0$ seems to be
  allowed. However, if we take into account that the spatial metric
  eq.~(\ref{eq:extremalmetric}) is odd in $\rho$, we can compensate the wrong
  sign in $e^{-U}$ with a change of sign in $\rho$. 
\end{enumerate}

In principle we have to consider the two aforementioned cases separately, but in the end
both can be written in a unified way, with the harmonic functions given by
\begin{equation}
  \label{eq:3}
  I_{0}  \ =\ \tfrac{1}{3} e^{-\sqrt{\frac{2}{3}}\phi_{\infty}} +|q_{0}|\rho 
  \hspace{.5cm},\hspace{.5cm}
  I_{1}  \ =\ \sigma \left\{
       \tfrac{2}{3} e^{\frac{1}{\sqrt{6}}\phi_{\infty}}+|q_{1}|\rho \right\}\, ,
\end{equation}

\noindent
and the mass and scalar charge are given by 
\begin{equation}
  \label{eq:4}
  M \ =\ |\mathcal{Z}_{(\sigma )}(\phi_{\infty},q)| 
  \hspace{.5cm}, \hspace{.5cm}
  \Sigma \ =\ 3 \partial_{\phi}\mathcal{Z}_{(\sigma )}(\phi_{\infty},q)\, .  
\end{equation}

Studying the near-horizon, {\em i.e.\/} $\rho\rightarrow \infty$, behavior we find that
\begin{align}
  \left. \phi_{\rm susy}\right|_{\rm h} & =\ -\sqrt{\tfrac{2}{3}} \log\left({\sigma \frac{2q_{0}}{q_{1}}}\right)\, ,\\
  & \nonumber \\
  \frac{A_{\rm h}}{2\pi^{2}} & = \sqrt{\tfrac{3^{3}}{2^{2}\, } |q_{0}| q_{1}^{2}}
    \, =\, \left[-V_{\rm bh}(\phi_{\rm h},q) \right]^{\frac{3}{4}} 
    \, =\, |\mathcal{Z}_{(\sigma )}(\phi_{\rm h},q)|^{\frac{3}{2}}\, .
\end{align}

These field configurations solve the same equations of motion all
values of $\sigma$, but they are only supersymmetric in the $\sigma$-branch of
the theory.
\subsection{Non-extremal solutions}
\label{sec-nonextsol}
The most general solution can be obtained by observing that the geodesic Lagrangian is separable:
by defining
\begin{equation}
  \label{eq:7}
  x \ \equiv\ U +\sqrt{\tfrac{2}{3}}\, \phi \;\;\; ,\;\;\;
  y \ \equiv\ U -\tfrac{1}{\sqrt{6}}\,\phi\, ,
\end{equation}

\noindent
the effective action eq.~(\ref{eq:effectived5}) takes the form
\begin{equation}
\mathcal{I}[x,y] \; =\; \int d\rho\
\left[\
\tfrac{1}{3}(\dot{x})^{2}
\ +\ \tfrac{2}{3}(\dot{y})^{2}
\ +\ 3 q^{2}_{0}e^{2x}
\ +\ \tfrac{3}{2} q^{2}_{1}e^{2y}\
\right]\; ,   
\end{equation}

\noindent
and its equations of motion can be integrated immediately. We do not need to
make any particular Ansatz, but should rather be able to recover it from the general
solution, which is\footnote{
  Please observe that this solution could also have been obtained by using the results
  obtained by Mohaupt {\&} Vaughan in ref.~\cite{Mohaupt:2010fk}.
}
\begin{eqnarray}
\label{eq:Effe10a}
e^{-3U} & =& \frac{3^{3}}{2^{2}}|q_{0}q_{1}^{2}| \
\left(\frac{\sinh{(B\rho+D)}}{B}\right)^{2}
\left(\frac{\sinh{(A\rho+C)}}{A}\right)\, ,  \\
& &\nonumber \\
\phi & =& -\sqrt{\tfrac{2}{3}} 
\log{
\left\{
\left|\frac{2q_{0}}{q_{1}}\right|
\left(\frac{B}{\sinh{(B\rho+D)}}\right)
\left(\frac{\sinh{(A\rho+C)}}{A}\right)
\right\}
}\, ,
\end{eqnarray}

\noindent
where $A,B,C$ and $D$ are (positive) integration constants. Their values are constrained
by the requirement of asymptotic flatness and related to the non-extremality
parameter $\mathcal{B}$ by the Hamiltonian constraint eq.~(\ref{eq:Effe2a})
\begin{equation}
\label{eq:Effe30}
             2B^{2}\ +\ A^{2}\; =\;  3 \mathcal{B}^{2}\, .  
\end{equation}

There are, then, three independent integration constants that must correspond
to the three independent physical parameters that are not the electric
charges: the mass $M$, the asymptotic value of the scalar
$\phi_{\infty}$ and the scalar charge $\Sigma$
(according to eq.~(\ref{eq:BBound}) $\mathcal{B}$ is a function of these three).
As the scalar charge is
not an attribute of point-like objects, we do not expect the
existence of regular black holes with $\Sigma\neq 0$ (scalar hair). However,
we know that regular black holes with $\Sigma \neq 0$ exist when $\Sigma$ is a
function of the other physical parameters $\Sigma_{0}(M,q,\phi_{\infty})$ (see {\em e.g.\/}
the supersymmetric case studied in the previous section). This kind of hair
is known as \textit{secondary hair} \cite{kn:CPW}, while $\Delta \Sigma\equiv
\Sigma-\Sigma_{0}$ is called \textit{primary hair} and its presence is generically
associated to singularities. 
\par
In order to make contact with Galli {\em et al.\/}'s Ansatz, we rewrite eqs.~(\ref{eq:Effe10a}) as
\begin{eqnarray}
\label{eq:Effe20a}
e^{-U} & =& e^{-U_{\rm susy}(\hat{I})}\ e^{(A+2B)\rho /3}\, ,\\
& \nonumber \\  
\label{eq:Effe20b} 
\phi & =&  \phi_{\rm susy}(\hat{I}) -\sqrt{\tfrac{2}{3}}(B-A)\rho\, ,
\end{eqnarray}

\noindent
where $e^{-U_{\rm susy}(I)}$ is given in eqs.~(\ref{eq:USUSY}) and
\begin{equation}
\phi_{\rm susy}(I) \ =\  
-\sqrt{\tfrac{2}{3}} \log{\left(\frac{2I_{0}}{I_{1}}\right)}\, ,
\end{equation}

\noindent
so there is no distinction between the branches.  The hatted ``harmonic''
functions are given by
\begin{align}
\hat{I}_{0} & = 
\tfrac{1}{3} e^{-\sqrt{\frac{2}{3}}\phi_{\infty}}
(2A)^{-1}
\left\{
\left(A+M+\sqrt{\tfrac{2}{3}}\, \Sigma \right)
 +
\left(A-M-\sqrt{\tfrac{2}{3}}\, \Sigma \right)
e^{-2A\rho} 
\right\}\, , 
\\
& \nonumber \\  
\hat{I}_{1} & = 
\tfrac{2}{3} e^{\frac{1}{\sqrt{6}}\phi_{\infty}}
(2B)^{-1}\left\{
\left(B+M-\tfrac{1}{\sqrt{6}}\,\Sigma \right)
 +
\left(B-M+\tfrac{1}{\sqrt{6}}\Sigma \right)
e^{-2B\rho} 
\right\}\, , 
\end{align}

\noindent
and the constants $A$ and $B$ are given by, taking the positive roots,
\begin{align}
A & = 
\sqrt{
\left(M+\sqrt{\tfrac{2}{3}}\, \Sigma \right)^{2}
-3^{2} q^{2}_{0}e^{2\sqrt{\frac{2}{3}}\phi_{\infty}}
}\, ,
\\
& \nonumber \\  
B & = 
\sqrt{
\left(M-\tfrac{1}{\sqrt{6}}\, \Sigma \right)^{2}
-\tfrac{3^{2}}{2^{2}} q^{2}_{1}e^{-\sqrt{\frac{2}{3}}\phi_{\infty}}
}\, .
\end{align}

A necessary condition for the solutions to become a product spacetime in the $\rho\rightarrow +\infty$
limit, thus signaling the occurrence of a horizon, can be read off from eq.~(\ref{eq:Effe20a}): $A+2B=\mathcal{B}$.  
This constraint together with the Hamiltonian constraint (\ref{eq:Effe30}) implies not only $A=B=\mathcal{B} \equiv \mathcal{B}_{0}$,
but also $\Sigma=\Sigma_{0}$ with\footnote{
  Only one of the solutions of
  the second degree equation for $\Sigma_{0}$ is valid, {\em i.e.\/} gives 
  rise to regular black holes.
}
\begin{equation}
\Sigma_{0} = -\sqrt{6}
\left\{ 
M -\sqrt{M^{2} +
3q^{2}_{0}e^{2\sqrt{\frac{2}{3}}\phi}
-\tfrac{3}{4}q^{2}_{1}e^{-\sqrt{\frac{2}{3}}\phi}
}
\right\}\, .  
\end{equation}

In that case, the form of the non-extremal solution is the one 
proposed by Galli {\em et al.\/} as a deformation of the supersymmetric one. 
In what follows we will only consider the regular solutions with no primary
scalar hair $\Sigma=\Sigma_{0}$, $\mathcal{B}=\mathcal{B}_{0}$. It is
useful to express these constants in terms of the asymptotic values of the
central charges of the two branches of the supersymmetric theory
$\mathcal{Z}_{(+)}$ and $\mathcal{Z}_{(-)}$:
\begin{equation}
  \label{eq:B0}
  \Sigma_{0} \ =\ -\sqrt{6} \left\{ M -\sqrt{C}\right\} 
  \hspace{.5cm}\mbox{and}\hspace{.5cm}
  \mathcal{B}^{2}_{0} \ =\ 5M^{2} 
      -3\mathcal{Z}_{(+)\, \infty}\tilde{\mathcal{Z}}_{(-)\, \infty} -4M\sqrt{C} \, .
\end{equation}

\noindent
where 
\begin{equation}
C 
\equiv 
M^{2} +\tfrac{3}{16}
\left(
3\mathcal{Z}^{2}_{(+)\, \infty}
+3\mathcal{Z}^{2}_{(-)\, \infty}
+10\mathcal{Z}_{(+)\, \infty}
\mathcal{Z}_{(-)\, \infty}
\right)\, .  
\end{equation}

\noindent
Further conditions for regularity of the bh's are the reality and
positivity of $\mathcal{B}^{2}_{0}$, which is the case if
\begin{equation}
\label{eq:extremalitybounds}
M^{2}\geq \mathcal{Z}^{2}_{(+)\, \infty}  
\hspace{1cm}\mbox{and}\hspace{1cm}
M^{2}\geq \mathcal{Z}^{2}_{(-)\, \infty}\, .
\end{equation}

\noindent
$\mathcal{B}_{0}$ vanishes only when one of the bounds is saturated, so there are in a given $\sigma$-branch two
extremal limits: one is supersymmetric and the other non-supersymmetric.

At the horizon, the scalar goes to the finite, yet $\phi_{\infty}$-dependent value 
\begin{equation}
\phi_{\rm h} \ =\ \phi_{\infty} -\sqrt{\tfrac{2}{3}}
\log{\left(
\frac{\mathcal{B}_{0}-M+2\sqrt{C}}{\mathcal{B}_{0}+2M-\sqrt{C}}
\right)} \, .
\end{equation}

The area of the horizon is easily found to be 
\begin{equation}
\frac{A_{\rm h}}{2\pi^{2}} \; =\; 
\sqrt{\left(\mathcal{B}_{0}-M+2\sqrt{C}\right)
\left(\mathcal{B}_{0}+2M-\sqrt{C}\right)^{2}}\, ,  
\end{equation}

\noindent
and the entropy can be computed from eq.~(\ref{eq:SSmarr}) $S=A_{\rm h}/3\pi$.
Also, using eq.~(\ref{eq:SSmarr}) the temperature is just
$T=\mathcal{B}_{0}/(3S)$ and vanishes in the extremal limits.

Let us end this section with a quick word on the extremal solutions: as
we found in the previous section the general family of non-extremal
solutions has two extremal limits, namely one given by $M=|\mathcal{Z}_{(+)\, \infty}|$ and the other one by
$M=|\mathcal{Z}_{(-)\, \infty}|$; the supersymmetry properties of the limiting solution will depend on the choice
of branch. In order to study them we have to take into account that
when one of the extremality bounds eq.~(\ref{eq:extremalitybounds}) is
saturated, the other one still holds. In other words: if (the absolute values
of) the two supercharges are different, the first bound that becomes saturated when we vary
the mass, will correspond to that of the largest supercharge. Which
supercharge is largest depends on the signs of the charges:
\begin{equation}
  \begin{array}{rclcrcl}
s_{0} & = & s_{1}\, ,\,\,\,\, & \Rightarrow  & |\mathcal{Z}_{(+)\, \infty}| &
\geq & |\mathcal{Z}_{(-)\, \infty}|\, ,\\
& & & & & & \\
s_{0} & = & -s_{1}\, ,\,\,\,\, & \Rightarrow  & |\mathcal{Z}_{(-)\, \infty}| &
\geq & |\mathcal{Z}_{(+)\, \infty}|\, .\\
  \end{array}
\end{equation}
As in the 4-dimensional examples studied in ref.~\cite{Galli:2011fq}, the
values of the charges determine completely the extremal limit. Taking this
into account is easy to see that we recover the extremal solutions found
before, whose supersymmetry properties depend on our choice of branch.
\section{Conclusions}
\label{sec-conclusions}

In this paper we have studied the generalization of the formalism of Ferrara,
Gibbons and Kallosh \cite{Ferrara:1997tw} to higher dimensions and we have
applied it to the construction of the non-extremal solutions of a simple model
of $N=2,d=5$ supergravity with just one modulus to check a proposal for a
generalization of the Ansatz of \cite{Galli:2011fq} to higher
dimensions. 

Instead of using this Ansatz directly, we have been able to integrate directly
the effective equations of motion of the model and we have found a general
solution with an independent scalar charge parameter $\Sigma$. Only when
$\Sigma$ is related to the mass, electric charges and asymptotic value of the
scalar by a given formula $\Sigma=\Sigma_{0}(M,q_{0},q_{1},\phi_{\infty})$ the
solutions are regular, {\em i.e.\/} black hole solutions and not naked singularities.
We can interpret these regular solutions as not having
\textit{primary scalar hair} in the sense of ref.~\cite{kn:CPW} and their form
fits perfectly in \cite{Galli:2011fq}'s Ansatz. 

Only a few examples of general families of solutions including singular
solutions with and regular solutions without primary scalar hair are known
\cite{kn:ALC2}. Most of the solutions known have only secondary hair: their
scalar charges are related to the masses, charges, and asymptotic values of
the moduli by certain expressions. In the supersymmetric cases these
expressions are related to the asymptotic values of the derivatives of the
central charges (or to the matter central charges\footnote{
  This follows from
  the general relation eq.~(\ref{eq:BBound}) for $\mathcal{B}=0$ plus the BPS
  bound $M=|\mathcal{Z}|$ and the expression of the black-hole potential in
  terms of the central charges.
} 
but in the general case it is not known how
to determine them before finding the explicit solutions. This is an important
problem for which no solution has been proposed. 

Here we have dealt with an extremely simple model. It is clear that to confirm
(or refute) the validity of \cite{Galli:2011fq}'s Ansatz more examples need to be studied. Work
in this direction is in progress.
\section*{Acknowledgments}
This work has been supported in part by the Spanish Ministry of Science and
Education grant FPA2009-07692, a Ram\'on y Cajal fellowship RYC-2009-05014,
the Princip\'au d'Asturies grant IB09-069, the Comunidad de Madrid grant
HEPHACOS S2009ESP-1473, and the Spanish Consolider-Ingenio 2010 program CPAN
CSD2007-00042. TO wishes to thank M.M.~Fern\'andez for her unfaltering
support.
\appendix

\end{document}